\documentclass[aps,prd,notitlepage,nofootinbib,preprintnumbers,superscriptaddress,10pt]{revtex4-2}
\usepackage{here}
\usepackage{graphicx}
\usepackage{amsmath,amsthm,amssymb}
\usepackage{bm}
\usepackage{color}

\usepackage{amsfonts}
\usepackage{dcolumn}
\usepackage{hyperref}
\allowdisplaybreaks[1]
\usepackage{stackengine}

\newcommand{\be}{\begin{eqnarray}}
\newcommand{\ee}{\end{eqnarray}}

\newcommand{\bem}{\begin{bmatrix}}
\newcommand{\eem}{\end{bmatrix}}

\newcommand{\mF}{\mathcal{F}}

\newcommand{\Mp}{M_{\rm p}}

\allowdisplaybreaks
\begin{document}
\title{Stealth black hole solutions in higher-order Maxwell-Einstein theories}

\author{Masato Minamitsuji}
\affiliation{Faculty of Health Sciences, Butsuryo College of Osaka, Sakai, 593-8328, Osaka, Japan}

\begin{abstract}
We study static and spherically symmetric black hole solutions in higher-order Maxwell-Einstein theories. We do not particularly focus on the degenerate classes of theories. For several specific choices of the coupling functions, we show that in the presence of the ordinary Maxwell kinetic term the  Reissner-Nordstr\"om-(anti-)de Sitter solution in the pure Maxwell-Einstein theory can also be a solution in generic classes of higher-order Maxwell-Einstein theories, and in the absence of the ordinary Maxwell kinetic term the Schwarzschild-(anti-)de Sitter solution with the nonzero electric field can be obtained. This corresponds to a stealth black hole solution as the electric field does not affect the spacetime geometry. We then focus on several degenerate classes of higher-order Maxwell-Einstein theories, and find that the dyonic Reissner-Nordstr\"om-(anti-)de Sitter solution in the pure Maxwell-Einstein theory can be a solution.
\end{abstract}

\date{\today}

\maketitle

%%%%%%%%%%%%%%%%%%%%%%%%%%%%%%%%%%%%%%%%%%
\section{Introduction}
\label{introsec}

According to the Lovelock theorem~\cite{Lovelock:1972vz}, General Relativity is the unique gravitational theory in four dimensions which contains two tensorial degrees of freedom and preserves the diffeomorphism invariance.
Whereas General Relativity has passed all the local and astrophysical tests~\cite{Will:2014kxa}, the standard $\Lambda$CDM model based on General Relativity has been plagued by tensions of today's observations~\cite{Riess:2019cxk,DiValentino:2021izs}.
With the improved accuracies, the tensions of the $\Lambda$CDM model with the observational data has become more significant.
This has motivated us to study alternative gravitational theories other than General Relativity~\cite{Sotiriou:2008rp,Clifton:2011jh,Will:2014kxa,Berti:2015itd}.

Scalar-tensor theories~\cite{Fujii:2003pa} are the representative candidates for alternative theories of gravitation.
Among them, Horndeski theories \cite{Horndeski:1974wa,Deffayet:2009wt,Kobayashi:2011nu} are the most general scalar-tensor theories which possess at most the second-order equations of motion despite the existence of the higher-derivative interactions of the scalar field. 
Moreover, degenerate higher-order scalar-tensor (DHOST) theories Refs.~\cite{Langlois:2015cwa,Crisostomi:2016czh,BenAchour:2016fzp} have been constructed by imposing the degeneracy of the higher-derivative equations of motion.
In DHOST theories, although the Euler-Lagrange equations contain higher-derivative terms, because of the degeneracy among them there are no Ostrogradsky ghosts.
Whereas in DHOST theories the Euler-Lagrange equations are degenerate in any choice of the gauge, in the U-DHOST theories \cite{DeFelice:2018ewo,DeFelice:2021hps} where the Euler-Lagrange equations are degenerate only in the unitary gauge there is also no appearance of the extra degrees of freedom except for the shadowy modes whose Euler-Lagrange equations are of the three-dimensional elliptic equations of motion.
Further extensions of higher-derivative scalar-tensor theories without the appearance of the Ostrogradsky ghosts have been extensively argued in e.g., Refs.~\cite{Langlois:2018dxi,Kobayashi:2019hrl,Babichev:2019twf,Gao:2019twq,Babichev:2021bim,Takahashi:2021ttd,Takahashi:2022mew,Naruko:2022vuh,Takahashi:2023vva,Hu:2024hzo,Babichev:2024eoh}.

Strongly gravitating compact objects such as black holes and neutron stars have been recognized as the most interesting and important research arenas for distinguishing General Relativity from the other theories of gravitation from both the theoretical and observational perspectives.
In many theories of gravitation, no-hair theorems for the static or stationary black hole solutions hold, and hence the external spacetime geometry around a black hole can be characterized only by the three conserved quantities as in General Relativity, namely, mass, angular momentum, and electric charge \cite{Israel:1967wq,Carter:1971zc,Chase,Bekenstein:1972ny,Hui:2012qt,Graham:2014ina,Graham:2014mda,Faraoni:2017ock}.
In Horndeski theories and scalar-tensor theories obtained as their extensions, various nontrivial black hole solutions have been obtained by relaxing at least one of the assumptions of the no-hair theorems~\cite{Babichev:2016rlq,Silva:2016smx}.
In the shift-symmetric higher-derivative scalar-tensor theories, the linear time dependence of the scalar field admits the stealth Schwarzschild or the stealth Schwarzschild-(anti-)-de Sitter solutions~\cite{Babichev:2013cya,Kobayashi:2014eva,Babichev:2015rva,Babichev:2016fbg,Babichev:2017guv,deRham:2019gha,Motohashi:2019ymr,Takahashi:2020hso,Mukohyama:2022enj,DeFelice:2022qaz}.
The existence of stealth black hole solutions was originally pointed out in Ref. \cite{Mukohyama:2005rw}.

The shift-symmetric Horndeski theories have been straightforwardly promoted to the vector-tensor theories with at most the second-order equations of motion.
The theories are called Generalized Proca theories~\cite{DeRham:2014wnv,DeFelice:2016yws,DeFelice:2016cri}.
Along the extension from Horndeski to DHOST theories, Generalized Proca theories have been generalized to Extended Vector-Tensor theories \cite{Kimura:2016rzw} where the highest order time derivative terms in the Euler-Lagrange are degenerate and hence there are no Ostrogradsky ghosts.
Further extensions of Generalized Proca and Extended Vector-Tensor theories have been argued in Refs.~\cite{GallegoCadavid:2019zke,GallegoCadavid:2020dho,GallegoCadavid:2021ljh,deRham:2021efp,Aoki:2021wew}.
Black hole solutions with nontrivial profiles of the vector field in Generalized Proca and Extended Vector-Tensor theories have been investigated in Refs.~\cite{Geng:2015kvs,Chagoya:2016aar,Minamitsuji:2016ydr,Cisterna:2016nwq,Fan:2016jnz,Heisenberg:2017xda,Heisenberg:2017hwb,Minamitsuji:2017aan,Babichev:2017rti,Chagoya:2017ojn,Fan:2017bka,Minamitsuji:2021gcq,Aoki:2023bmz}.
The stealth Schwarzschild solutions and the Schwarzschild-(anti-)de Sitter solutions in shift symmetric Horndeski theories have been straightforwardly promoted to those in Generalized Proca and Extended Vector-Tensor theories~\cite{Chagoya:2016aar,Minamitsuji:2016ydr,Fan:2016jnz,Heisenberg:2017xda,Heisenberg:2017hwb,Minamitsuji:2017aan,Babichev:2017rti,Minamitsuji:2021gcq,Aoki:2023bmz}.
Besides the straightforward promotion from the shift-symmetric Horndeski theories, there also exist the stealth Schwarzschild solutions and the Schwarzschild- (anti-)de Sitter solutions with the nonzero electric charge for a certain choice of the coupling parameter \cite{Chagoya:2016aar,Minamitsuji:2016ydr,Heisenberg:2017xda,Heisenberg:2017hwb,Minamitsuji:2017aan,Minamitsuji:2021gcq}.

Whereas Generalized Proca and Extended Vector-Tensor theories explicitly break the $U(1)$ invariance in the sector of the vector field, there had been no formulation of degenerate vector-tensor theories preserving the $U(1)$ symmetry which forbids the appearance of the Ostrogradsky ghosts.
Recently, a general framework of higher-derivative vector-tensor theories preserving the $U(1)$ symmetry have been constructed in Ref.~\cite{Colleaux:2023cqu},
and the degeneracy conditions in the Minkowski and general curved spacetimes have been systematically constructed in Refs.~\cite{Colleaux:2024ndy,Colleaux:2025vtm}, respectively. 
The starting action in Ref.~\cite{Colleaux:2023cqu} contains the scalar products of the tensors which are constructed from the higher-order powers of the  metric, the electromagnetic field strength and its dual tensor (See Eq.~\eqref{dual}) with the Riemann tensor and those with the quadratic order product of the first order derivative of the electromagnetic field strength tensor.
Using the symmetries/anti-symmetries of the Riemann and the electromagnetic field strength tensors and the identities associated with these tensors, building blocks for the $U(1)$-invariant interactions with the higher derivatives have been identified.
The resultant action contains totally 21 higher-derivative interactions, i.e., 3 products including the Riemann tensor and 18 products including the quadratic order product of the first-order derivative of the electromagnetic field strength tensor.~\cite{Colleaux:2023cqu}.
On the higher-order Maxwell theories in the Minkowski spacetime which are of the quadratic order in the first-order derivatives of the electromagnetic field strength tensor, the Hamiltonian analysis has been presented in Ref. ~\cite{Colleaux:2024ndy}.
It has been shown that only the degeneracy of the kinetic matrix is not sufficient to eliminate all the Ostrogradsky ghosts. 
In Ref. \cite{Colleaux:2025vtm}, the analysis of the degeneracy conditions in higher-order Maxwell-Einstein theories has been extended to curved spacetimes and the classification of the degenerate higher-order Maxwell-Einstein theories have presented.

In this article, we investigate the static and spherically symmetric black hole solutions in higher-order Maxwell-Einstein theories.
We focus on both the nondegenerate and degenerate theories.
The article is divided into two parts.
In the first part, we focus on theories before imposing the degeneracy conditions constructed in Ref.~\cite{Colleaux:2023cqu}.
Starting with the static and spherically symmetric ansatz for the metric and the vector field, we derive the conditions under which higher-order Maxwell-Einstein theories possess the Reissner-Nordstr\"om and Schwarzschild solutions in the pure Maxwell-Einstein theory.
We consider the following two cases:
In the case that the action contains the ordinary Maxwell kinetic, we derive the conditions under which the Reissner-Nordstr\"om solution as in the ordinary Maxwell-Einstein theory is also a solution in higher-order Maxwell-Einstein theories. 
On the other hand, in the case  that the action does not contain the ordinary Maxwell kinetic term, we derive the conditions under which the Schwarzschild solution exists. 
The Schwarzschild solution contains the nonzero electric charge which does not affect the spacetime geometry, and corresponds to a stealth black hole solution.
%%%
In the second part, we focus on the existence of the dyonic Reissner-Nordstr\"om-(anti-)de Sitter solution in several classes of the degenerate higher-order Maxwell-Einstein theories.
We show that only Class ${\rm C}_1 {\rm II}$ of the degenerate high-order Maxwell theories obtained in Ref.  \cite{Colleaux:2025vtm} admits the dyonic  Reissner-Nordstr\"om-(anti-)de Sitter solution. 
For such a solution, we also find that in the absence of the ordinary Maxwell kinetic term the black hole solution turns to the dyonic Schwarzschild-de Sitter solution which corresponds to a stealth solution.
For other classes of degenerate theories, we confirm that there are no such solutions.

The article is constructed as follows:
In Sec. \ref{sec2}, we review higher-order Maxwell-Einstein theories, introduce the static and spherically symmetric ansatz, and obtain the Euler-Lagrange equations.
In Sec.~\ref{sec3}, we derive the conditions for the existence of the Reissner-Nordstr\"om solution in higher-order Maxwell-Einstein theories where the coefficients in front of all the higher-order Maxwell-Einstein interactions are constant. 
We also derive the conditions for the existence of the Schwarzschild solution in higher-order Maxwell-Einstein theories where the coefficients in front of all the higher-order Maxwell-Einstein interactions are constant. 
The solution can be regarded as a stealth black hole solution, as the electric charge does not affect the spacetime geometry.
We also analyze  the existence of the Reissner-Nordst\"om-(anti)de Sitter solution and the Schwarzschild-(anti-)de Sitter solutions in the same class of higher-order Maxwell-Einstein theories with the cosmological constant.
In Secs.~\ref{sec4} and \ref{sec5}, we perform the same analysis for higher-order Maxwell-Einstein theories which are of the quartic and sixth orders in the electromagnetic field strength and its first-order derivative, respectively.
In Sec.~\ref{sec6}, we focus on several classes of degenerate higher-order Maxwell-Einstein theories, and argue the existence of the dyonic Reissner-Nordstr\"om-(anti-)de Sitter solution and the dyonic Schwarzschild- (ant-)de Sitter solution.
In Sec. \ref{sec7}, we close the article after giving a brief summary and conclusion.

%%%%%%%%%%%%%%%%%%%%%%%%%%%%%%%%%%%%%%%%%%
\section{Higher-order Maxwell-Einstein theories}
\label{sec2}

\subsection{Action}

Ref.~\cite{Colleaux:2023cqu} constructed the general framework of $U(1)$-invariant higher-derivative vector-tensor theories.
The starting action is given by 
\be
\label{action0}
S
=
\int 
d^4 x
{\cal L}
:=
\int d^4x \sqrt{-g}
\left[
L_0
+
\frac{1}{4}
A^{abcd}R_{abcd}
+
B^{abc,def}
\nabla_a F_{bc}
\nabla_d F_{ef}
\right],
\ee
where the indices $a,b,c,\cdots$ run the four-dimensional spacetime, the tensor $g_{ab}$ represents the spacetime metric, $g$ is the determinant of $g_{ab}$, i.e., $g:={\rm det}(g_{ab})$, $\nabla_a$ represents the covariant derivative associated with the metric $g_{ab}$, $R_{abcd}$ represents the Riemann curvature tensor associated with $g_{ab}$, $F_{ab}:=\nabla_a A_b-\nabla_b A_a=\partial_a A_b-\partial_b A_a$ represents the electromagnetic field strength, respectively.
The Lagrangian function $L_0$, and the rank-4 and rank-6 tensors $A^{abcd}$ and $B^{abc,def}$ are constituted by the metric $g_{ab}$, the electromagnetic field strength tensor $F_{ab}$, and the dual field strength tensor to 
\be
\label{dual}
\left({\ast F}\right)_{ab}:=\frac{1}{2}
\varepsilon_{abcd}F^{cd},
\ee
with $\varepsilon^{abcd}$ being the Levi-Civita tensor.
Eq.~\eqref{action0} corresponds to the most general action which is constituted by the scalar products including the Riemann tensor and the quadratic-order product of the first-order derivative of the electromagnetic field strength.

Using the symmetries/anti-symmetries of the tensors $R_{abcd}$ and $F_{ab}$ and the identities associated with the derivatives of them, the form of $A^{abcd}$ and $B^{abc,def}$ is reduced in Ref.~\cite{Colleaux:2023cqu}.
The resultant action~\cite{Colleaux:2023cqu} is given by 
\be
S
&=&
\int 
d^4 x
{\cal L}
=
\int d^4x
\sqrt{-g}
\left\{
L_0
+
\sum_{i=1}^3
a_i
{\cal R}_i
+
\sum_{i=1}^{18}
b_i
{\cal F}_i
\right\},
\label{action}
\ee
where the scalar products are given by 
\be
\label{ffr}
{\cal R}_1=R,
\qquad 
{\cal R}_2
=F^{ab}F^{cd}R_{abcd},
\qquad 
{\cal R}_3
=F_2^{ab}R_{ab},
\ee
and 
\be
\label{weight0}
\mF_1
&=&
\nabla^b F_{ab}
\nabla^c F_{c}{}^a,
\\
\mF_2
&=&
F_2^{ec}
\nabla_e F_{ab}
\nabla^b F_c{}^a,
\qquad
\mF_3
=
F_2^{ad}
\nabla^c F_{ab}
\nabla^b F_{cd},
\qquad
\mF_4
=
F^{eb}_2
\nabla_e
F_{ab}
\nabla_c F^{ca},
\nonumber
\\
\mF_5
&=&
F_2^{ad}
\nabla^b 
F_{ab}
\nabla^c
F_{cd},
\qquad 
\mF_6
=
F^{ad}
F^{ec}
\nabla_e
F_{ab}
\nabla^b
F_{cd},
\nonumber\\
\mF_7
&=&
F^{eb}
F^{cf}
\nabla_e F_{ab}
\nabla_f F_c{}^a,
\quad 
\mF_8
=
F^{ad}
F^{eb}
\nabla_e F_{ab}
\nabla^c F_{cd},
\label{weight2}
\\
\mF_9
&=&
F_2^{ab}
F^{eb}
\nabla_e F_{ab}
\nabla^c F_{cd},
\label{weight3}
\\
\mF_{10}
&=&
F_3^{ab}
F^{ec}
\nabla_e F_{ab}
\nabla^b F_{cd},
\qquad 
\mF_{11}
=
F_3^{ad}
F^{eb}
\nabla_e 
F_{ab}
\nabla^c
F_{cd},
\qquad 
\mF_{12}
=
F_2^{eb}
F_2^{cf}
\nabla_e
F_{ab}
\nabla_f
F_c{}^a,
\nonumber\\
\mF_{13}
&=&
F_2^{ad}
F_2^{eb}
\nabla_e F_{ab}
\nabla^c
F_{cd},
\qquad
\mF_{14}
=
F^{ad}
F^{eb}
F_2^{cf}
\nabla_e  F_{ab}
\nabla_f F_{cd},
\qquad 
\mF_{15}
=
F_2^{ad}
F^{eb}
F^{cf}
\nabla_{e}
F_{ab}
\nabla_f
F_{cd},
\label{weight4}
\\
\mF_{16}
&=&
F_3^{ad}
F_2^{eb}
\nabla_e F_{ab}
\nabla^c F_{cd},
\qquad 
\mF_{17}
=
F_2^{ad}
F^{eb}
F_2^{cf}
\nabla_e F_{ab}
\nabla_f F_{cd},
\label{weight5}
\\
\mF_{18}
&=&
F_3^{ad}
F^{eb}
F_2^{cf}
\nabla_e F_{ab}
\nabla_f F_{cd}.
\label{weight6}
\ee
Here, $R_{ab}=g^{cd}R_{acbd}$ represents the Ricci tensor associated with the metric $g_{ab}$, $R=g^{ac}g^{bd}R_{abcd}$ is the Ricci scalar associated with $g_{ab}$, $F_{0}^a{}_b:=\delta^a{}_b$, $F_1^a{}_b:=F^{a}{}_b$, $F_2^a{}_b:=F^{a}{}_c F^c{}_b$, $F_3^a{}_b:=F^{a}{}_c F^c{}_d F^d{}_b$, and in general for $I\geq 2$
\be
F_I{}^a{}_b
:=
F^a{}_{c_1} 
F^{c_1}{}_{c_2} 
F^{c_2}{}_{c_3}
\cdots 
F^{c_{I-2}}{}_{c_{I-1}}
F^{c_{I-1}}{}_b.
\label{FIs}
\ee
Using the Cayley-Hamilton theorem, $F_4{}^a{}_b$ can be written in terms of $F_0{}^a{}_b$ and $F_2{}^a{}_b$ with the coefficients depending on $F_2$ and $F_4$, where
\be
F_I
:=g_{ab} F_I^{ab}
=
F^{c_I}{}_{c_1} 
F^{c_1}{}_{c_2} 
F^{c_2}{}_{c_3}
\cdots 
F^{c_{I-2}}{}_{c_{I-1}}
F^{c_{I-1}}{}_{c_I}.
\ee
Thus, $F_I{}^a{}_b$ with even indices $I=6,8,\cdots$ can be written as a linear combination of $F_0{}^a{}_b$ and $F_2{}^a{}_b$ with the coefficients depending on 
$F_2$ and $F_4$, and similarly $F_I{}^a{}_b$ with odd indices $I=5,7,\cdots$ can be written as a linear combination of $F_1{}^a{}_b$ and $F_3{}^a{}_b$ with the coefficients depending on 
$F_2$ and $F_4$.
Thus, any higher-order scalar products of $F_I{}^{a}{}$ can be written in terms of $F_2$ and $F_4$~\cite{Colleaux:2023cqu}.
Moreover, since the scalar product 
\be
{\ast F}_2:=\left(\ast F\right)_{ab}F^{ab},
\ee
can also be written in terms of $F_2$ and $F_4$ as 
\be
\label{24star}
\left({\ast F}_2\right)^2
=4\left(
F_4-\frac{1}{2}F_2^2\right),
\ee
any higher-order scalar products of $F_I{}^{a}{}$ can be written in term of $F_2$ and ${\ast F}_2$.
The coefficients $a_i$($i=1,2,3$) and $b_i$($i=1,2,3,\cdots,17,18$) are in general the functions of $F_2$ and $F_4$ \cite{Colleaux:2023cqu}, or using the identity \eqref{24star}, they can instead be regarded as the functions of $F_2$ and ${\ast F}_2$.
The Lagrangians in Eqs.~\eqref{action0} and \eqref{action} include three types of terms: (i) standard Maxwell and cosmological constant terms, (ii) explicit couplings to the Riemann tensor, and (iii) terms quadratic in first derivatives of the EM field strength, i.e., involving second derivatives of $F_a$.
Although covariant derivatives acting on a vector field produce curvature via their commutator, the derivative-squared terms $\mathcal{F}_i$ in Eq.~\eqref{action}) are constructed solely from contractions of $\nabla_aF_{bc}$ and are manifestly $U(1)$-gauge invariant. They are not expressed through commutators and thus do not double-count curvature effects already present in the $a_2$ and $a_3$ terms.
Even when $a_2 = a_3 = 0$, the $b_i \mathcal{F}_i$ terms do not implicitly introduce curvature couplings with the $U(1)$ symmetry. They arise purely from second derivatives of the vector field, and their structure ensures linear independence from the explicit curvature terms in the effective theory.

We note that the theory \eqref{action} is not free from the instabilities associated with the Ostrogradsky ghosts, and hence further conditions have to be imposed to eliminate them.
Ref.~\cite{Colleaux:2024ndy} investigates higher-order Maxwell theories in the Minkowski spacetime constituted by the scalar products including the quadratic power of the first-order derivative of the electromagnetic field strength tensor, and shows that the degeneracy of the kinetic matrix for the second-order time derivatives in the Lagrangian is not sufficient to eliminate all the Ostrogradsky ghosts and additional conditions need to be imposed.
%%%%%%%
Ref.~\cite{Colleaux:2025vtm} investigates the degeneracy conditions for higher-order Maxwell-Einstein theories in general curved spacetimes \eqref{action} with Eqs. \eqref{ffr}-\eqref{weight6}.
In Sec. \ref{sec5}, we focus on the concrete classes of the degenerate higher-order Maxwell-Einstein theories~\cite{Colleaux:2025vtm}.

Until Sec.~\ref{sec4}, we focus on the most general action of higher-order Maxwell-Einstein theories \eqref{action} with Eqs. \eqref{ffr}-\eqref{weight6} derived in Ref. \cite{Colleaux:2023cqu}.
Throughout this article, we assume that the coefficient in front of the Einstein-Hilbert term is given by 
\be
a_1=\frac{\Mp^2}{2},
\ee
where $\Mp$ denotes the reduced Planck mass, and the Lagrangian $L_0$ is given by the ordinary Maxwell kinetic term and the cosmological constant
\be
\label{L0}
L_0=c_1 F_2-\Mp^2\Lambda,
\ee
where $\Lambda$ represents the cosmological constant.
We also assume that $c_1>0$.
We note that although the canonically normalized Maxwell kinetic term corresponds to $c_1=\frac{1}{4}$, we keep $c_1$ as a free parameter.

\subsection{Static and spherically symmetric spacetime}

We assume the static and spherically symmetric spacetime and the vector field which, respectively, are given by 
\be
\label{metric}
ds^2=g_{ab}dx^a dx^b
     &=& -q^2f(r) dt^2+\frac{dr^2}{h(r)}+r^2\left(d\theta^2+\sin^2\theta d\varphi^2\right),
\ee
where $t$, $r$, and $(\theta,\varphi)$ are the temporal, radial, and angular coordinates, respectively, and $f(r)$, $h(r)$ and $A_0(r)$ are functions of $r$.
$f(r)$ and $h(r)$ represent the temporal and radial components of the spacetime metric, respectively.
Moreover, we assume that the vector field  is purely electric, i.e., 
\be
\label{vector}
A_a dx^a
&=&
q A_0(r)dt,
\ee
where $A_0(r)$ represents the temporal component of the vector field.
With the ansatz \eqref{vector}, only the nontrivial component of the electromagnetic field strength is given by $F_{tr}=-\partial_ r A_0(r)=-A_0'(r)$.
In Eqs. \eqref{metric} and \eqref{vector}, the constant $q$ represents the time reparametrization symmetry, and without loss of generality we may set $q=1$ by the redefinition of the time coordinate $t'=\sqrt{q}t$.
Substituting the ansatz~\eqref{metric} and \eqref{vector} into the action \eqref{action} and varying it with respect to $f(r)$, $h(r)$, and $A_0(r)$, taking the highest order derivative of $r$ into consideration, we obtain the Euler-Lagrange equations which are schematically given by 
\be
\label{ELeq}
{\cal E}_f
&:=&
\frac{\partial {\cal L}}{\partial f(r)}
-
\frac{\partial}{\partial r}
\left(
\frac{\partial {\cal L}}{\partial f'(r)}
\right)
+
\frac{\partial^2}{\partial r^2}
\left(
\frac{\partial {\cal L}}{\partial f''(r)}
\right),
\nonumber\\
{\cal E}_h
&:=&
\frac{\partial {\cal L}}{\partial h(r)}
-
\frac{\partial}{\partial r}
\left(
\frac{\partial {\cal L}}{\partial f'(r)}
\right),
\nonumber\\
{\cal E}_{A_0}
&:=&
-
\frac{\partial}{\partial r}
\left(
\frac{\partial {\cal L}}{\partial A_0'(r)}
\right)
+
\frac{\partial^2}{\partial r^2}
\left(
\frac{\partial {\cal L}}{\partial A_0''(r)}
\right).
\ee
In Eq. \eqref{ELeq}, the absence of $\frac{\partial {\cal L}}{\partial A_0(r)}$ is because of the $U(1)$ symmetry.

%%%%%%%%%%%%
\section{Reissner-Nordstr\"om and Schwarzschild solutions in higher-order Maxwell-Einstein theories with constant coefficients}
\label{sec3}

In this section, we assume that in the action \eqref{action} the coefficients are constants, i.e.,
\be
\label{constant_choice}
a_2={\rm constant}, \qquad a_3={\rm constant}, \qquad b_i={\rm constant}  \,\,\,(i=1,2,3,\cdots 17,18).
\ee

\subsection{Reissner-Nordstr\"om solution}
\label{sec3a}

First, we assume the presence of the ordinary Maxwell kinetic term and the vanishing cosmological constant
\be
\label{choice1}
c_1\neq 0,
\qquad \Lambda=0,
\ee
in Eq.~\eqref{L0}, which allow asymptotically flat black hole solutions.
The pure Maxwell-Einstein theory corresponds to the choice of
\be
\label{EM_action}
a_2&=&a_3=0,
\qquad 
b_i=0\quad (i=1,2,\cdots,17,18),
\ee
where for the ansatz of the metric \eqref{metric} the vector field \eqref{vector2} the unique vacuum static and spherically symmetric solution is given by the Reissner-Nordstr\"om solution
\be
\label{exactRN}
h(r)&=&1-\frac{2M}{r}+\frac{2c_1 Q^2}{\Mp^2 r^2},
\nonumber\\
f(r)&=&h(r),
\nonumber\\
A_0(r)&=&\frac{Q}{r},
\ee
where the constants $M$ and $Q$ represent the mass and electric charge of the black hole, respectively.

We find that even in the presence of higher-order Maxwell-Einstein interactions the Reissner-Nordstr\"om solution \eqref{exactRN} is a solution in the higher-order Maxwell-Einstein theories \eqref{action} with Eq. \eqref{constant_choice}, in the case that the constant coefficients in front of higher-order Maxwell-Einstein theories satisfy the following conditions
\be
\label{cond_coef}
a_2&=&a_3=0,
\nonumber\\
b_7&=&b_2-\frac{3}{2}b_3+b_6,
\qquad 
b_8=b_3-b_4,
\nonumber\\
b_{13}&=&- b_{11},
\qquad 
b_{15}=b_{10}-b_{12}-b_{14},
\nonumber\\
b_{18}&=&0,
\ee
Whereas all other coefficients $b_1$, $b_2$, $b_3$, $b_4$, $b_5$, $b_6$, $b_9$, $b_{10}$, $b_{11}$, $b_{12}$, $b_{14}$, $b_{16}$, and $b_{17}$ remain arbitrary.
Thus, for the choice of Eq. \eqref{constant_choice}, the existence of the Reissner-Nordstr\"om solution \eqref{exactRN} requires the absence of the derivative interactions of the electromagnetic field strength to the Riemann tensor.

We confirm this by making use of the expansion in the large distance limit $r\to \infty$.
In the case of $c_1\neq 0$, the Euler-Lagrange equations \eqref{ELeq} yield the asymptotic solution in the large distance region
\be
\label{large_distance}
f(r)
&=&
1+
\sum_{n=1}^\infty
\frac{f_n}{r^n},
\qquad 
h(r)
=
1+
\sum_{n=1}^\infty
\frac{h_n}{r^n},
\qquad 
A_0(r)
=
\sum_{n=1}^\infty
\frac{d_n}{r^n},
\ee
where the leading-order coefficients are uniquely given
\be
\label{expansion1}
f_1
&=&
-2M,
\quad 
f_2
=
\frac{2c_1 Q^2}{\Mp^2},
\quad 
f_3=0,
\quad 
f_4
=
-\frac{2a_2Q^2}{\Mp^2},
\quad 
f_5
=
\frac{2(a_2+a_3)MQ^2}
{\Mp^2},
\nonumber\\
f_6
&=&
-
\frac{4(a_2+4a_3)c_1Q^4}
    {5\Mp^4},
\quad
f_7
=
-\frac{160a_2^2 M Q^2}
      {7c_1\Mp^2},
\nonumber\\
f_8
&=&
\frac{Q^2}{14c_1\Mp^4}
\nonumber\\
&\times&
\Big[
3c_1
\left(
-8a_3^2
+
\left(
 2b_2
+3b_3
-2b_6
+2b_7
\right)
\Mp^2
\right)
Q^2
+
28 a_2 a_3
\left(
8M^2\Mp^2-5c_1 Q^2
\right)
+
4a_2^2
\left(
64M^2\Mp^2+265c_1Q^2
\right)
\Big],
\\
h_1
&=&
-2M,
\quad 
h_2
=
\frac{2c_1Q^2}{\Mp^2}
\quad 
h_3
=0,
\quad 
h_4
=
\frac{2(-4a_2+3a_3)Q^2}
    {\Mp^2},
\quad
h_5
=
\frac{2(7a_2-5a_3)MQ^2}
     {\Mp^2},
\nonumber\\
h_6
&=&
\frac{4\left(-16a_2+11a_3\right)c_1Q^4}
{5\Mp^4},
\quad 
h_7
=
-
\frac{32\left(5a_2-3a_3\right)a_2 MQ^2}
{c_1 \Mp^2},
\nonumber\\
h_8
&=&
\frac{2Q^2}{7c_1\Mp^4}
\nonumber\\
&\times&
\Big[
c_1
\left(
274a_3^2
-\left(
2b_2+3b_3-2b_6+2b_7\right)\Mp^2
\right)
Q^2
+
128a_2^2
\left(
8M^2\Mp^2+17c_1Q^2
\right)
-
56
a_2a_3
\left(
11M^2\Mp^2+31 c_1 Q^2
\right)
\Big],
\\
d_1
&=&
Q,
\quad 
d_2=d_3=0,
\quad 
d_4
=
\frac{a_2MQ}{c_1},
\quad
d_5
=
-
\frac{\left(9a_2+a_3\right)Q^3} 
      {5\Mp^2},
\quad 
d_6=\frac{6a_2b_1MQ}
        {c_1^2},
\nonumber\\
d_7
&=&
\frac{Q}{14c_1^2\Mp^2}
\nonumber\\
&\times&
\Big[
c_1
\Big(
-24a_3^2
+56a_3b_1
+
\big(
6b_2
+2b_3
+7b_4
-6b_6
+6b_7
+7b_8
\big)
\Mp^2
\Big)Q^2
-168
a_2b_1
\big(
M^2\Mp^2
+
2c_1Q^2
\big)
\nonumber\\
&+&
16a_2^2
\big(
2M^2\Mp^2+5c_1Q^2
\big)
\Big],
\nonumber\\
d_8
&=&
\frac{MQ}{14c_1^3}
\Big[
1260 a_2 b_1^2
-7
\left(
2b_2+b_3
+
2
\left(
b_4-b_6+b_7+b_8
\right)
\right)
c_1^2Q^2
\nonumber\\
&-&
\frac{4c_1^2Q^2}
    {\Mp^2}
\left(
54a_2^2
+
21a_2(a_3-10b_1)
-
14a_3(a_3-2b_1)
\right)
\Big].
\ee
We omit to show the higher-order terms as they are quite lengthy.
Imposing the conditions \eqref{cond_coef} yields
\be
h_{i\geq 3}=0,
\quad 
f_{i\geq 3}=0,
\quad 
a_{i\geq 2}=0,
\ee
showing the existence of the Reissner-Nordstr\"om solution \eqref{exactRN}.

%%%%%%%%%%%%%%
\subsection{Stealth Schwarzschild solution}
\label{sec3b}

Second, in the same theory \eqref{action} with Eq.~\eqref{constant_choice},
we assume the absence of both the ordinary Maxwell kinetic term and the cosmological constant 
\be
\label{choice2}
c_1= 0,
\qquad 
\Lambda=0,
\ee
in Eq.~\eqref{L0}.
We also assume that the ansatz of the metric and vector field is given by Eqs. \eqref{metric} and \eqref{vector}, respectively.

We find that the Schwarzschild solution with the nonzero electric charge
\be
\label{Sch}
h(r)&=&1-\frac{2M}{r},
\nonumber\\
f(r)&=&h(r),
\nonumber\\
A_0(r)&=&\frac{Q}{r},
\ee
where $M$ and $Q$ represent the mass and electric charge of the black hole, respectively is also a solution in higher-order Maxwell-Einstein theories \eqref{action} with Eq. \eqref{constant_choice}, respectively, in the case that the constant coefficients satisfy the conditions \eqref{cond_coef}, whereas all other coefficients $b_1$, $b_2$, $b_3$, $b_4$, $b_5$, $b_6$, $b_9$, $b_{10}$, $b_{11}$, $b_{12}$, $b_{14}$, $b_{16}$, and $b_{17}$ remain arbitrary.
Since the electric field does not affect the spacetime metric functions, the solution corresponds to a stealth black hole solution.
The existence of the stealth black hole is a common feature in higher-derivative scalar-tensor and vector-tensor theories
~\cite{Mukohyama:2005rw,Babichev:2013cya,Kobayashi:2014eva,Babichev:2015rva,Minamitsuji:2016ydr,Babichev:2016fbg,Chagoya:2016aar,Babichev:2017guv,Minamitsuji:2017aan,Heisenberg:2017xda,deRham:2019gha,Motohashi:2019ymr,Heisenberg:2017hwb,Takahashi:2020hso,Minamitsuji:2021gcq,Mukohyama:2022enj,DeFelice:2022qaz}.

In the case of Eq. \eqref{choice2}, in the large distance limit $r\to\infty$, the Euler-Lagrange equations \eqref{ELeq} yield Eq.~\eqref{large_distance}, where the leading-order coefficients are uniquely given by 
\be
\label{expansion2}
f_1
&=&
-2M,
\quad 
f_2=f_3=0,
\quad 
f_4
=
-\frac{2a_2Q^4}{\Mp^2},
\nonumber\\
f_5
&=&
\frac{
2MQ^2\left(24a_2^2+5a_2b_1+5a_3b_1\right)}
{5b_1\Mp^2},
\quad 
f_6=
-
\frac{8a_2M^2Q^2
\left(
12a_2^2-a_2b_1+5a_3 b_1
\right)
}{5b_1^2\Mp^2},
\nonumber\\
f_7
&=&
\frac{4a_2M^3Q^2}
     {315b_1^3\Mp^2}
\left(
 1760a_2^3
-1768a_2^2b_1
-630a_3b_1^2
+81a_2b_1\left(14a_3+5b_1\right)
\right),
\\
h_1
&=&
-2M,
\quad 
h_2
=0,
\quad 
h_3
=0,
\quad 
h_4
=
\frac{2Q^2(-4a_2+3a_3)}{\Mp^2},
\nonumber\\
h_5
&=&
\frac{2\left(24a_2^2-16a_2a_3+7a_2b_1-5a_3b_1\right)MQ^2}
    {b_1\Mp^2},
\quad 
h_6
=
-\frac{4a_2M^2Q^2}{5b_1^2\Mp^2}
\left[
144a_2^2-90a_2a_3
+29a_2b_1-20a_3b_1
\right],
\nonumber\\
h_7
&=&
\frac{4a_2M^3Q^2}{45b_1^3\Mp^2}
\left(
1760a_2^3
+3a_2(146a_3-45b_1)b_1
+
90 a_3b_1^2
-8a_2^2
(132a_3+95b_1)
\right),
\\
d_1&=&Q,
\quad 
d_2=-\frac{a_2MQ}{b_1},
\quad 
d_3=\frac{2a_2(4a_2-5b_1)M^2Q}{15b_1^2r^3},
\nonumber\\
d_4
&=&
-\frac{2a_2M^3Q}{45b_1^3}
\left(
4a_2^2-17a_2b_1+15b_1^2
\right),
\nonumber\\
d_5
&=&
-\frac{Q}{6300b_1^4\Mp^2}
\Bigg[
-256a_2^4M^4\Mp^2
+2432 a_2^3 b_1M^4\Mp^2
\nonumber\\
&+&
45b_1^3
\Big(
-24a_3^2+84a_3b_1
+
\left(
6b_2+2b_3+7b_4-6b_6+6b_7+7b_8
\right)
\Mp^2
\Big)
Q^2
\nonumber\\
&&
-1260a_2b_1^3
\left(
-4M^4\Mp^2+3b_1Q^2
\right)
+
48a_2^2b_1^2
\left(
-139M^4\Mp^2
+75 b_1 Q^2
\right)
\Bigg].
\ee
We omit to show the higher-order terms as they become quite lengthy.
Imposing the conditions~\eqref{cond_coef} yield
\be
h_{i\geq 2}=0,
\quad 
f_{i\geq 2}=0,
\quad 
a_{i\geq 2}=0,
\ee
showing the existence of the Schwarzschild solution \eqref{Sch}. 
In the theories \eqref{action} with Eq. \eqref{constant_choice}, Eq.~\eqref{cond_coef} indicates the existence of the stealth Schwarzschild solution does not admits the derivative interaction of the Riemann tensor $A^{abcd}R_{abcd}$ in the original action \eqref{action0}.

\subsection{Reissner-Nordstr\"om-(anti-)de Sitter solution}
\label{sec3c}

Third, it is straightforward to extend the solutions to the asymptotically de Sitter or anti-de Sitter solutions by adding the nonzero cosmological constant $\Lambda$ to $L_0$ as Eq.~\eqref{L0}.
In the presence of the ordinary Maxwell kinetic term and the nonvanishing cosmological constant
\be
\label{choice3}
c_1\neq 0,
\qquad \Lambda\neq0,
\ee
in Eq. \eqref{L0}, the Reissner-Nordstr\"om-(anti-)de Sitter solution
\be
\label{exactRNdS}
h(r)&=&
-\frac{\Lambda}{3}r^2
+
1-\frac{2M}{r}+\frac{2c_1 Q^2}{\Mp^2 r^2},
\nonumber\\
f(r)&=&h(r),
\nonumber\\
A_0(r)&=&\frac{Q}{r},
\ee
corresponds to the unique static and spherically symmetric solution in 
the pure Maxwell-Einstein theory with the choice \eqref{EM_action}.
For the ansatz of the metric \eqref{metric} and the vector field \eqref{vector}, we find that even in higher-order Maxwell-Einstein theories \eqref{action} with the constant coefficients \eqref{constant_choice}, the Reissner-Nordstr\"om-(anti-)de Sitter solution \eqref{exactRNdS} can also be a solution in the case that the constant coefficients in Eq.~\eqref{action} satisfy the algebraic conditions \eqref{cond_coef}.

\subsection{Stealth Schwarzschild-de Sitter solution}
\label{sec3d}

Fourth, in the absence of the ordinary Maxwell kinetic term and the nonvanishing cosmological constant 
\be
\label{choice4}
c_1= 0,
\qquad 
\Lambda\neq 0,
\ee
in Eq.~\eqref{L0}, for the same ansatz as Eqs. \eqref{metric} and \eqref{vector}, we find that the Schwarzschild-de Sitter solution with the nonzero electric charge
\be
\label{SchdS}
h(r)&=&
-\frac{\Lambda}{3}r^2
+
1-\frac{2M}{r},
\nonumber\\
f(r)&=&h(r),
\nonumber\\
A_0(r)&=&\frac{Q}{r},
\ee
where $M$ and $Q$ correspond to the mass and electric charge of the black hole, respectively, is also a solution in higher-order Maxwell-Einstein theories \eqref{action} with the constant coefficients \eqref{constant_choice}, in the case that the coefficients satisfy the conditions \eqref{cond_coef}.
Since the electric field does not affect the spacetime geometry, the solution \eqref{SchdS} corresponds to a stealth black hole solution.
This is the straightforward extension of the stealth Schwarzschild solution discussed in subsection \ref{sec3b}.

%%%%%%%%%%%%
\section{Solutions in higher-order Maxwell-Einstein theories with the quartic order interactions in the electromagnetic field strength tensor}\label{sec4}

In this section, we choose the coefficients in the action \eqref{action} as 
\be
\label{quartic_higher}
a_2={\bar a}_2 F_2,
\qquad 
a_3={\bar a}_3 F_2,
\qquad 
b_1={\bar b}_1 F_2,
\ee
where ${\bar a}_1$, ${\bar a}_2$, and ${\bar b}_1$ are constants, and $b_i$($i=2,3,4,5,6,7,8$) as nonzero constants.
We also assume that $b_i$($i=9,10,\cdots,18$) are set to $0$.
In the action \eqref{action} with Eq.~\eqref{quartic_higher}, the higher-order Maxwell-Einstein terms are of the quartic order in the electromagnetic field strength tensor $F_{ab}$ and its first-order derivative.
As defined in Eq.~\eqref{FIs}, $F_2^{ab}$ and $F_3^{ab}$ are of quadratic and cubic order in the electromagnetic field strength tensor $F_{ab}$, respectively. Thus, the terms from ${\cal F}_1$ to ${\cal F}_8$ are at most of quartic order in electromagnetic field strength tensor $F_{ab}$, whereas the terms from ${\cal F}_{9}$ to ${\cal F}_{18}$ are of higher order, i.e., beyond quintic order in the electromagnetic field strength tensor $F_{ab}$. In this section, we focus on derivative interactions involving the electromagnetic field strength tensor $F_{ab}$ up to quartic order in $F_{ab}$.

\subsection{Reissner-Nordstr\"om solution}

First, we assume the presence of the ordinary Maxwell kinetic term and the vanishing cosmological constant, Eq. \eqref{choice1} in Eq. \eqref{L0}.
We also assume the ansatz of the metric \eqref{metric} and the vector field \eqref{vector}.
We then find that the Reissner-Nordstr\"om solution \eqref{exactRN} is also a solution in the higher-order Maxwell-Einstein theories \eqref{action} with Eq. \eqref{quartic_higher} in the case that the coefficients satisfy the following conditions
\be
\label{cond_coef3}
{\bar a}_3&=&2{\bar a}_2,
\nonumber\\
b_7&=&14{\bar a}_2-b_2-\frac{3}{2}b_3+b_6,
\qquad 
b_8=-12{\bar a}_2+b_3-b_4,
\ee
whereas all other coefficients ${\bar b}_1$, $b_2$, $b_3$, $b_4$, $b_5$, and $b_{6}$ remain arbitrary.

%%%%%%%%%%%%%%
\subsection{Stealth Schwarzschild solution}

Second, we assume the absence of the ordinary Maxwell kinetic term and the vanishing cosmological constant, Eq.~\eqref{choice2} in Eq.~\eqref{L0}.
We also assume that the ansatz of the metric and vector field is given by Eqs. \eqref{metric} and \eqref{vector}, respectively.
We then find that the Schwarzschild solution with the nonzero electric charge \eqref{Sch} can be obtained in the higher-order Maxwell-Einstein theories \eqref{action} with Eq. \eqref{quartic_higher} in the case that the coefficients satisfy the conditions \eqref{cond_coef3}.
This Schwarzschild solution can be regarded as a stealth black hole solution.
The difference from the theory discussed in Sec. \ref{sec3} is that the existence of the stealth Schwarzschild solution admits the nonzero higher-order coupling to the Riemann tensor so long as the coefficients satisfy $a_3=2a_2$.

\subsection{Reissner-Nordstr\"om-(anti-)de Sitter solution}

Third, in the presence of the ordinary Maxwell kinetic term and the nonvanishing cosmological constant, Eq. \eqref{choice3} in Eq. \eqref{L0}, for the ansatz of the metric \eqref{metric} and the vector field \eqref{vector}, the Reissner-Nordstr\"om-(anti-)de Sitter solution \eqref{exactRNdS} can also be a solution in the higher-order Maxwell-Einstein theories in the case that the coefficients in the action \eqref{action} with Eq. \eqref{quartic_higher} satisfy the conditions \eqref{cond_coef3}.

\subsection{Stealth Schwarzschild-(anti-)de Sitter solution}

Fourth, in the absence of the ordinary Maxwell kinetic term and the nonvanishing cosmological constant \eqref{choice4} in Eq. \eqref{L0}, we also assume the same ansatz of the metric and vector field as Eqs. \eqref{metric} and \eqref{vector}.
We find that the Schwarzschild-(anti-)de Sitter solution with the nonzero electric charge \eqref{SchdS} can be obtained in the higher-order Maxwell-Einstein theories \eqref{action} with Eq. \eqref{sixth_interaction} in the case that the coefficients satisfy the conditions  \eqref{cond_coef3}.
Since the electric field does not affect the spacetime geometry, the solution can be regarded as a stealth black hole solution.
The difference from the theory discussed in Sec. \ref{sec3} is that the existence of the stealth Schwarzschild solution admits the nonzero higher-order coupling to the Riemann tensor so long as the coefficients satisfy $a_3=2a_2$.

%%%%%%%%%%%%
\section{Solutions in higher-order Maxwell-Einstein theories with the sixth order interactions in the electromagnetic field strength tensor}
\label{sec5}

In this section, we choose the coefficients in the action \eqref{action} as 
\be
\label{sixth_interaction}
a_2&=&{\tilde a}_2 F_4,
\qquad 
a_3={\tilde a}_3 F_4,
\qquad 
b_1={\tilde b}_1 F_4,
\nonumber\\
b_i&=&{\tilde b}_i F_2\qquad (i=2,3,\cdots,8),
\ee
where ${\tilde a}_1$, ${\tilde a}_2$, ${\tilde b}_1$, and ${\tilde b}_i$ ($i=2,3,\cdots,8$) are constants, and $b_i$($i=10,11,\cdots, 15$) are nonzero constants.
We also assume that $b_i$($i=9,16,17,18$) are set to $0$.
All the higher-order Maxwell-Einstein terms in the action \eqref{action} with Eq. \eqref{sixth_interaction} are of the sixth-order in the electromagnetic field strength or its first-order derivative.

\subsection{Reissner-Nordstr\"om solution}

First, we assume the presence of the ordinary Maxwell kinetic term and the vanishing cosmological constant, Eq. \eqref{choice1} in Eq.~\eqref{L0}.
For the ansatz of the metric \eqref{metric}, we find that the Reissner-Nordstr\"om solution \eqref{exactRN} is a solution in the higher-order Maxwell-Einstein theories \eqref{action} with Eq. \eqref{sixth_interaction} in the case that the coefficients in the action \eqref{action} with Eq.~\eqref{sixth_interaction} satisfy the following conditions
\be
\label{cond_coef5}
{\tilde a}_3&=&2{\tilde a}_2,
\nonumber\\
b_{13}&=&-20{\tilde a}_2-b_{11}+2{\tilde b}_3-2{\tilde b}_4-2{\tilde b}_8,
\qquad 
b_{15}=22 {\tilde a}_2+b_{10}-b_{12}-b_{14}-2{\tilde b}_2-3{\tilde b}_3+2{\tilde b}_6-2{\tilde b}_7,
\ee
whereas all other coefficients ${\tilde b}_1$, ${\tilde b}_2$, ${\tilde b}_3$, ${\tilde b}_4$, ${\tilde b}_5$, ${\tilde b}_{6}$, ${\tilde b}_7$, ${\tilde b}_8$, $b_{10}$, $b_{11}$, $b_{12}$, and $b_{14}$ remain arbitrary.

%%%%%%%%%%%%%%
\subsection{Stealth Schwarzschild solution}

Second, we assume the absence of the ordinary Maxwell kinetic term and the vanishing cosmological constant, Eq. \eqref{choice2} in Eq. \eqref{L0}.
We also assume the same ansatz of the metric and vector field as Eqs. \eqref{metric} and \eqref{vector}, respectively.
We find that the Schwarzschild solution with the nonzero electric charge \eqref{Sch} can be obtained in the higher-order Maxwell-Einstein theories \eqref{action} with Eq. \eqref{sixth_interaction} in the case that the coefficients satisfy the conditions \eqref{cond_coef5}.
This Schwarzschild solution can be regarded as a stealth black hole solution.
The difference from the theory discussed in Sec. \ref{sec3} is that the existence of the stealth Schwarzschild solution admits the nonzero higher-order coupling to the Riemann tensor so long as the coefficients satisfy 
$a_3=2a_2$.

\subsection{Reissner-Nordstr\"om-(anti-)de Sitter solution}

Third, in the presence of the ordinary Maxwell kinetic term and the nonvanishing cosmological constant, Eq. \eqref{choice3} in Eq.~\eqref{L0}, for the ansatz of the metric \eqref{metric} and the vector field \eqref{vector}, we find that the Reissner-Nordstr\"om-(anti-)de Sitter solution \eqref{exactRNdS} is also a solution in the higher-order Maxwell-Einstein theories \eqref{action} with Eq.~\eqref{sixth_interaction} in the case that the coefficients in the action \eqref{action} satisfy the conditions \eqref{cond_coef5}.

\subsection{Stealth Schwarzschild-de Sitter solution}

Fourth, in the absence of the ordinary Maxwell kinetic term and the nonvanishing cosmological constant, Eq. \eqref{choice4} in Eq. \eqref{L0}, we find that the Schwarzschild-(anti-) de Sitter solution with the nonzero electric charge \eqref{SchdS} is a solution in the higher-order Maxwell-Einstein theories \eqref{action} with Eq.~\eqref{sixth_interaction}, in the case that the coefficients satisfy the conditions \eqref{cond_coef5}.
Since the electric field does not affect the spacetime geometry, the solution corresponds to a stealth black hole solution.
The difference from the theory discussed in Sec. \ref{sec3} is that the existence of the stealth Schwarzschild solution admits the nonzero higher-order coupling to the Riemann tensor so long as the coefficients satisfy 
$a_3=2a_2$.

%%%%%%%%%%%%%%%
\section{Dyonic black holes in degenerate higher-order Maxwell-Einstein theory}
\label{sec6}

In this section, we focus on the degenerate classes of higher-order Maxwell-Einstein theories discussed in Ref. \cite{Colleaux:2025vtm}.
The kinetic matrix for the first-order time derivative terms of the metric and the second-order time derivative terms of the vector field in the action \eqref{action} is schematically given by 
\be
\label{kineticmatrix}
{\cal M}
=
\begin{pmatrix}
{\cal E} & {\cal D} \\
{\cal D} & {\cal C} \\
\end{pmatrix},
\ee
where the components of the ${\cal E}$ and ${\cal C}$ represent the coefficients for the quadratic-order terms of the second-order time derivatives of the vector field and the first-order time derivatives of the metric functions in the Lagrangian, respectively, and ${\cal D}$ represents the coefficients for the products of the second-order time derivative of the vector field and the first-order time derivative of the metric functions.
Imposing the degeneracy conditions on Eq. \eqref{kineticmatrix}, 
several classes of the degenerate higher-order Maxwell-Einstein theories are obtained in Ref.~\cite{Colleaux:2025vtm}.

In this section, we focus on the static and spherically symmetric metric ansatz \eqref{metric}. 
For the vector field, in addition to the purely electric ansatz \eqref{vector}, we also consider the dyonic ansatz
\be
\label{vector2}
A_a dx^a= A_0(r)dt-P\cos\theta d\varphi,
\ee
where $P$ represents the magnetic charge.
In the presence of both nonzero electric and magnetic charges, 
both $F_{ab}$ $\left(\ast F\right)_{ab}$ can be nontrivial.

\subsection{Totally degenerate theories}

As discussed in Refs. \cite{Colleaux:2025vtm}, the case where ${\cal D}={\cal E}=0$, the Euler-Lagrange are at most of the second-order for the metric and of the third-order for the vector field. 
Among them, we focus on the class of the totally degenerate case ${\cal C}={\cal D}={\cal E}=0$, higher-order Maxwell-Einstein interactions reduce to the form of $\alpha_0({\cal F},{\cal G})\left(\ast F\right)^{ab}\nabla_a {\cal F}\nabla_b{\cal G}$, where we define
\be
{\cal F}:=\frac{1}{4}F_2,
\qquad 
{\cal G}:=\frac{1}{4} \left({\ast F}_2\right),
\ee
and $\alpha_0({\cal F},{\cal G})$ represents a free function of ${\cal F}$ and ${\cal G}$.
In this subsection, we consider a subclass of the degenerate higher-order Maxwell-Einstein theories where the Einstein-Hilbert term and the ordinary Maxwell kinetic term are added, which is given by
\be
\label{action30}
S=\int d^4x
\sqrt{-g}
\left[
\frac{\Mp^2}{2}
\left(
R-2\Lambda
\right)
+
c_1 F_2
+
\alpha_0
{\cal F}^p{\cal G}^q
\left(\ast F\right)^{cd}\nabla_c {\cal F}\nabla_d {\cal G}
\right],
\ee
where $\alpha_0$, $p\geq 0$, and $q\geq 0$ are constants.
$\alpha_0$ represents the strength of the derivative interaction of the electromagnetic field strength tensor.

Assuming the static and spherically symmetric spacetime metric ansatz \eqref{metric}, the purely electric ansatz for the vector field \eqref{vector} automatically satisfies $\left(\ast F\right)^{ab}=0$ and hence ${\cal G}=0$.
Thus, the higher-order Maxwell-Einstein interactions proportional to $\alpha_0$ trivially vanish, and the Reissiner-Nordstr\"om solution-(anti-) de Sitter solution in the pure Maxwell-Einstein theory \eqref{exactRN} is automatically a solution of Eq.~\eqref{action30}.

On the other hand, the dyonic Reissner-Nordst\"rom-(anti-) de Sitter solution in the pure Maxwell-Einstein theory, which is explicitly given by
\be
\label{dyonic}
f(r)
&=&
h(r)
=
-\frac{\Lambda}{3}r^2
+1
-\frac{2M}{r}
+
\frac{2c_1\left(P^2+Q^2\right)}{\Mp^2r^2},
\nonumber\\
A_0(r)&=&\frac{Q}{r},
\ee
is also a solution in the theory \eqref{action30} with any finite value of $\alpha_0$, in the case that 
\be
P=Q,
\ee
namely, in the case that the magnetic charge $P$ is equal to the electric charge $Q$, except for the case of $p=0$ where there is no dyonic Reissiner-Nordstr\"om solution-(anti-)de Sitter solution given by Eq. \eqref{dyonic} for any relation between $P$ and $Q$.

Degenerate classes of higher-order Maxwell-Einstein theories are also obtained for the case of ${\cal E}={\cal D}=0$ and ${\cal C}\neq 0$ in Ref. \cite{Colleaux:2025vtm}. 
However, we do not focus on these classes in this article.

\subsection{Theories with degenerate derivative coupling to the Riemann tensor}

In this subsection, we focus on the subclass of the generic higher-order Maxwell-Einstein theories \eqref{action0}
\be
\label{action02}
S=\int d^4x \sqrt{-g}
\left[
L_0
+
\frac{1}{4}
A^{abcd}R_{abcd}
\right],
\ee
which is a class of Eq. \eqref{action0} with $ B^{abc,def}=0$, namely there is no quadratic order terms of the first-order derivatives of the electromagnetic field strength.
The degeneracy conditions for the class of higher-order Maxwell-Einstein theories \eqref{action02} are classified in Ref. \cite{Colleaux:2025vtm}.
Among them, we focus on Class ${\rm C}_1 {\rm II}$ of the degenerate higher-order Maxwell-Einstein theories, which is added to the pure Einstein-Hilbert term and the pure Maxwell kinetic term
\be
\label{action3}
S=\int d^4x
\sqrt{-g}
\left[
\frac{\Mp^2}{2}
\left(
R-2\Lambda
\right)
+
c_1 F_2
+
\beta
\left(
{\cal F},
{\cal G}
\right)
\left(
2{\cal F}
\left(
{\ast F}
\right)^{ab}
F^{cd}
+
{\cal G}
\left(
F^{ab}
F^{cd}
-
\left(
{\ast F}
\right)^{ab}
\left(
{\ast F}
\right)^{cd}
\right)
\right)
C_{acbd}
\right],
\ee
where $\beta \left({\cal F},{\cal G} \right)$ is an arbitrary function of ${\cal F}$ and ${\cal G}$,
\be
C_{abcd}
:=
R_{abcd}
-
\frac{1}{2}
\left(
g_{ac}R_{bd}
-
g_{ad}R_{bc}
+
g_{bd}R_{ac}
-
g_{bc}R_{ad}
\right)
+
\frac{1}{6}
R
\left(
g_{ac}g_{bd}
-
g_{ad}g_{bc}
\right),
\ee
represents the Weyl tensor associated with the metric $g_{ab}$.
For the static and spherically symmetric spacetime metric ansatz \eqref{metric}, the purely electric ansatz for the vector field \eqref{vector} automatically satisfies $\left(\ast F\right)^{ab}=0$ and hence ${\cal G}=0$.
Thus, the higher-order Maxwell-Einstein terms which are proportional to $\beta$ trivially vanish, and the Reissiner-Nordstr\"om solution-(anti-)de Sitter solution in the pure Maxwell-Einstein theory \eqref{exactRN} is automatically a solution in the given class of degenerate higher-order Maxwell-Einstein theories.

Moreover, for the static and spherically symmetric metric ansatz \eqref{metric} and the dyonic ansatz of the vector field \eqref{vector2}, we find the identical relation
\be
\left(
2{\cal F}
\left(
{\ast F}
\right)^{ab}
F^{cd}
+
{\cal G}
\left(
F^{ab}
F^{cd}
-
\left(
{\ast F}
\right)^{ab}
\left(
{\ast F}
\right)^{cd}
\right)
\right)
C_{acbd}
=0.
\ee
Hence, the dyonic Reissner-Nordstr\"om-(anti-)de Sitter solution in the pure Maxwell-Einstein theory, Eq. \eqref{dyonic}, is always a solution.
We emphasize that the existence of the dyonic solution does not depend on the form of the function $\beta \left({\cal F},{\cal G}\right)$ in the action \eqref{action3}.

In the case of $c_1= 0$ in Eq. \eqref{action3}, namely in the absence of the ordinary Maxwell-Einstein kinetic term,
there exists the dyonic Schwarzschild-(anti-)de Sitter black hole solution
\be
\label{dyonic2}
f(r)
&=&
h(r)
=
-\frac{\Lambda}{3}r^2
+1
-\frac{2M}{r},
\nonumber\\
A_0(r)&=&\frac{Q}{r}.
\ee
Since both the electric and magnetic charges do not affect the spacetime geometry, the solution \eqref{dyonic2} can be regarded as a stealth black hole.
The existence of an electrically charged dyonic stealth black hole solution is a unique feature in the degenerate higher-order Maxwell-Einstein theories.

We also examine the degenerate theories of Class ${\rm C}_3$ classified in Ref. \cite{Colleaux:2025vtm} and find that there are no dyonic Reissner-Nordstr\"om-(anti-)de Sitter solution and no stealth Schwarzschild-(anti-)de Sitter solution.
The inspection of the other classes of the degenerate higher-order Maxwell-Einstein theories are left for future works.

%%%%%%%%%%%%%%%%%%%%%%%%%%
\section{Conclusions}
\label{sec7}

In this article, we have investigated the static and spherically symmetric solutions in higher-order Maxwell-Einstein theories. 
The theories correspond to a higher-derivative extension of the pure Maxwell-Einstein theory with the $U(1)$ symmetry.
The action of higher-order Maxwell theories contains scalar products of the tensors constructed with the electromagnetic field strength tensor with the Riemann tensor, and those with the quartic order power of the first-order derivatives of the electromagnetic field strength.
We have focused on the existence of the Reissner-Nordstr\"om and Schwarzschild solutions and their extensions with the cosmological constants.
We have focused on several choices of the coupling functions.
First, we have considered the case where all the coefficients in front of higher-order Maxwell-Einstein interactions are constant.
Second, we have considered the cases where the higher-order Maxwell-Einstein terms are of the quartic and sixth orders in the electromagnetic field strength tensor and its first order derivative, respectively.

In both cases, we have shown that in the presence of the ordinary Maxwell kinetic term, in the case that the coefficients in front of the higher-order Maxwell-Einstein terms satisfy certain conditions, the Reissner-Nordstr\"om solution in the ordinary Maxwell-Einstein theory is also a solution in higher-order Maxwell-Einstein theories.
In this case, the higher-order interactions of the electromagnetic field strength tensor do not contribute to the modification of the black hole  solution. 
In the first case where all the coefficients in front of higher-order Maxwell-Einstein interactions are constant, the existence of the Reissner-Nordstr\"om solution requires the vanishing coefficients for the contractions with the Riemann tensor.
In the second case, the coefficients for the contractions with the Riemann tensor do not need to vanish, but have to satisfy the certain tuning relation.
By adding the cosmological constant, the Reissner-Nordstr\"om-(anti-)de Sitter solution in the ordinary Maxwell-Einstein theory is also a solution in higher-order Maxwell-Einstein theories where the coefficients in front of higher-order Maxwell-Einstein interactions satisfy the certain the same relations as those without the cosmological constant.

On the other hand, in both the cases, the absence of the ordinary Maxwell kinetic term leads to the Schwarzschild and Schwarzschild-(anti-)de Sitter solutions with nonzero electric charge, where the electric charge does not affect the spacetime geometry. 
The conditions for the existence of the Schwarzschild and Schwarzschild-(anti-)de Sitter solutions among the coefficients in front of the higher-order Maxwell-Einstein terms remain the same as those for the existence of the Reissner-Nordstr\"om and Reissner-Nordstr\"om-(anti-)de Sitter solutions.
Since the electric charge in the sector of the vector field does affect the spacetime geometry, these Schwarzschild and Schwarzschild-(anti-)de Sitter solutions can be regarded as stealth black hole solutions.

Finally, we have shown that the dyonic Reissner-Nordstr\"om-(anti-)de Sitter solution in the pure Maxwell-Einstein theory is also a solution in Class ${\rm C}_1 {\rm II}$ of degenerate higher-order Maxwell-Einstein theories discussed in Ref. \cite{Colleaux:2025vtm}.
The existence of these dyonic solutions does not depend on the coupling function associated with degenerate higher-order Maxwell-Einstein interactions, and there is no specific relation between the electric and magnetic charges.
We have also considered the totally degenerate class of the degenerate higher-order Maxwell-Einstein theories, where the dyonic Reissner-Nordstr\"om-(anti-)de Sitter solution is a solution so long as the magnetic charge is equal to the electric charge.
In some other classes of degenerate higher-order Maxwell-Einstein theories discussed in Ref.~ \cite{Colleaux:2025vtm}, we have found that the dyonic Reissner-Nordstr\"om-(anti-)de Sitter solution in the pure Maxwell-Einstein theory is not a solution.

We have not fully explored the Reissner-Nordstr\"om-(anti-)de Sitter and Schwarzschild-(anti-)de Sitter solutions in all the classes of the degenerate higher-order Maxwell-Einstein theories yet. This will be left for future works. 
It would also be of interest to investigate more nontrivial static and spherically symmetric black hole solutions carrying electric and magnetic charges, beyond the standard Reissner--Nordstr\"om--(anti-) de Sitter and Schwarzschild--(anti-)de Sitter solutions. The study of rotating black hole solutions represents a promising direction for further research. At this stage, however, we do not have conclusive results on these more general configurations. In the present work, we have restricted our analysis to Schwarzschild- or Reissner--Nordstr\"om-like metrics and have imposed corresponding forms of the $U(1)$ vector field from the outset. This simplification was adopted to ensure analytical tractability and to focus on a particular class of solutions. To explore the possibility of other types of black hole solutions with nontrivial hair, it would be necessary to relax these assumptions and adopt a more general ansatz for both the spacetime metric and the vector field. We expect that such an analysis would be significantly more complex and technically involved. For this reason, we have deferred a detailed investigation of these broader possibilities to future work. Nevertheless, the presence of the new $U(1)$-invariant higher-derivative interactions considered in this work may plausibly give rise to more nontrivial black hole solutions. We hope to return to these issues in a future publication.

%%%%%%%%%%%%%%%%%
\section*{ACKNOWLEDGMENTS}
This work was supported by the funds of Butsuryo College of Osaka.

\appendix

\bibliography{refs}
\end{document}